\documentclass[aps,amsmath,amssymb,prb,twocolumn,showpacs]{revtex4}
\usepackage{bm}
\usepackage{epsfig}

\newcommand{\cC}{{\mathcal{C}}}

\newcommand{\cJ}{{\mathcal{J}}}

\begin{document}

\preprint{draft}

\title{Delocalization in Coupled Luttinger Liquids with Impurities}

\author{Stefan Scheidl}
\author{Simon Bogner}
\author{Thorsten Emig}

\affiliation{Institut f\"ur Theoretische Physik, Universit\"at zu
K\"oln, Z\"ulpicher Stra\ss e 77, D-50937 K\"oln, Germany}

\date{\today}

\begin{abstract}
  We study effects of quenched disorder on coupled two-dimensional
  arrays of Luttinger liquids (LL) as a model for stripes in
  high-$T_c$ compounds. In the framework of a renormalization-group
  analysis, we find that weak inter-LL charge-density-wave couplings
  are always irrelevant as opposed to the pure system.  By varying
  either disorder strength, intra- or inter-LL interactions, the
  system can undergo a delocalization transition between an insulator
  and a novel strongly anisotropic metallic state with LL-like
  transport.  This state is characterized by short-ranged
  charge-density-wave order, the superconducting order is quasi
  long-ranged along the stripes and short-ranged in the transversal
  direction.
\end{abstract}


\maketitle

\section{Introduction}

Quasi-one-dimensional electron liquids play a paradigmatic role in
describing the conductive properties of a variety of physical systems
such as organic conductors,\cite{Jerome91} quantum-Hall
systems,\cite{vOppen00} and striped phases in high-$T_\textrm{c}$
compounds.\cite{Zaanen+89,Emery+97} Recent studies of weakly coupled
Luttinger liquids (LL) have provided evidence for the stability of
non-Fermi-liquid behavior in more than one dimension
\cite{Emery+00,Vishwanath+01,Mukhopadhyay+01} as opposed to results
for an isotropic 2D Fermi gas.\cite{Engelbrecht+90} This remarkable
result is a consequence of the combined effect of single
particle/Cooper pair tunneling and Coulomb interactions between the
LLs.  Earlier studies either excluded hopping,\cite{Lee+77} treated
all inter-LL interactions separately as weak perturbations
\cite{Klemm+76} or focused on single particle tunneling for strong
repulsive intra-LL interactions only.\cite{Wen90} In Refs.
\onlinecite{Emery+00,Vishwanath+01,Mukhopadhyay+01} it was shown that
backscattering and particle hopping processes between the LLs can be
irrelevant for sufficiently strong inter-LL forward scattering.  The
resulting state was called ``sliding Luttinger liquid'' (SLL).  For a
large range of interactions, these processes can be partially relevant
and lead to charge-density wave (CDW), transverse superconductor (SC)
or Fermi Liquid (FL)
phases.\cite{Emery+00,Vishwanath+01,Mukhopadhyay+01} Experiments have
provided evidence for 1D transport in high-$T_\textrm{c}$
compounds.\cite{Noda+99} Theoretically, novel and to date essentially
unexplored behavior can arise from disorder, which is induced by
doping in these materials.

Here, we examine the role of electron scattering by a random impurity
potential.  For a single LL it was shown
\cite{Apel82,Apel+82,Giamarchi+88} that a delocalization transition
can occur with increasing electron attraction and that repulsive
interactions always lead to localization.  On the other hand, for
coupled LLs, a simple scaling analysis suggests that disorder would be
irrelevant at least in the SLL phase.\cite{Mukhopadhyay+01} However,
using a renormalization-group (RG) analysis, we show that disorder
profoundly modifies the characteristic properties of these systems.
It turns out that a delocalization transition persists in analogy to
single LLs. Where Josephson inter-stripe couplings are irrelevant, the
delocalized phase can be identified with a new state of matter, which
we dub disordered stripe metal (DSM).  In contrast to the SLL state of
the pure system, even in this delocalized phase, there exists only
short ranged longitudinal CDW order due to impurity forward
scattering. Because of this scattering process, we also find a strong
tendency towards the destruction of transverse CDW order. Thus, the
novel DSM state combines short ranged CDW order and quasi long-ranged
longitudinal superconducting order with LL-like transport properties.
Interestingly, it has a much wider stability region in comparison to
the pure system's SLL state.

\section{General Model and Renormalization}

We assume a spin gap in the LLs, as present in stripes in
high-$T_\textrm{c}$ compounds.\cite{Emery+97} The low-energy charge
excitations of noninteracting stripes (labeled by $j$) can be
described\cite{note:Review} by the bosonic phase fields $\Phi_j$ and
their dual fields $\Theta_j$ with an action
\begin{equation}
  \label{S_j}
  S^0 = \frac{1}{2\pi} \sum_j \int_{x\tau} \Big[ v_J 
  (\partial_x \Theta_j)^2 + v_N (\partial_x \Phi_j)^2
  - 2i\partial_\tau \Phi_j \partial_x \Theta_j  \Big] .
\end{equation}
The characteristic velocities $v_N$ and $v_J$ include forward
scattering by intra-stripe \textit{interactions}, whereas backward
scattering is assumed to be irrelevant.

Following Giamarchi and Schulz,\cite{Giamarchi+88} forward and
backward scattering by weak \textit{impurities} (denoted by IFS and
IBS, respectively) is described in terms of the action
\begin{subequations}
\label{S.IS}
\begin{eqnarray}
  S^{\rm IFS} &=& -\frac{\sqrt{2}}{\pi}\sum_j\int_{x\tau} 
  \eta_j(x) \partial_x \Phi_j,
  \label{S.IFS}
  \\
  S^{\rm IBS} &=& \frac{1}{\pi \alpha} \sum_j \int_{x\tau} \left\{ \xi_j(x)
    e^{i(\sqrt{2}\Phi_j-2k_{\rm F}x)} + \text{h.c.} \right\} . \ \ 
  \label{S.IBS}
\end{eqnarray}
\end{subequations}
$\eta_j(x)$ and $\xi_j(x)$ are Gaussian random variables with zero
mean and correlations $\overline{\eta_i(x)\eta_j(x')}
=\frac{1}{2}D_\eta \delta_{ij}\delta(x-x')$ and
$\overline{\xi^*_i(x)\xi_j(x')}=D_\xi \delta_{ij}\delta(x-x')$.
$\alpha$ is the infinitesimal regularization length of bosonization.

In order to describe coupled arrays of stripes, we first include forward
scattering due to density-density interactions between the stripes,
such as screened Coulomb and electron-phonon interactions.  The
corresponding action reads
\begin{equation}
  \label{S0}
  S^V = \frac{1}{2\pi} \!\sum_{i\neq j}\! \int_{x\tau}\!\!
  \partial_x\Phi_i V_{i-j}\partial_x \Phi_j.
\end{equation}
In principle, analogous couplings between $\partial_x \Theta_j$ can be
added,\cite{Emery+00,Vishwanath+01,Mukhopadhyay+01} but are dropped
for simplicity. Their inclusion can be achieved by an obvious
generalization of our analysis, and will not modify our main results.
Since $S^V$ is bilinear in $\Phi_i$, it can be included in the
momentum-space representation of the action $S^0$ by replacing $v_N$
in Eq. (\ref{S_j}) by the $q_\perp$ dependent velocity $\tilde
v_N(q_\perp)=v_N+V_{q_\perp}$ where $V_{q_\perp}$ is the Fourier
transform of $V_i$ (we use the spacing between stripes as transverse
length unit). The restriction to low energy excitations underlying
bosonization obviously requires the stability condition
$v_N+V_{q_\perp}>0$.  The conventionally defined Luttinger parameter
$K=\sqrt{v_J/v_N}$ is then generalized to
\begin{equation}
  \label{Ktilde}
  \tilde K(q_\perp)=\sqrt{v_J/\tilde v_N(q_\perp)}=
  K/\sqrt{1+V_{q_\perp}/v_N}
\end{equation}
in order to include inter-stripe forward scattering.  

Besides these forward scattering processes, we allow also for pairwise
hopping between stripes, given by the transverse CDW and SC couplings
\begin{subequations}
\label{S-CDW-SC}
\begin{eqnarray}
  \label{S-CDW}
  S^{\rm CDW} &=& \sum_{i\neq j} {\cal C}_{i-j} \int_{x\tau} 
  \cos\left[\sqrt{2}(\Phi_i - \Phi_j)\right],
  \\
  \label{S-SC}
   S^{\rm SC} &=& \sum_{i\neq j} {\cal J}_{i-j} \int_{x\tau} 
  \cos\left[\sqrt{2}(\Theta_i - \Theta_j)\right].
\end{eqnarray}
\end{subequations}
Since we assume the presence of a spin gap, single electron hopping is
irrelevant and can be ignored. \cite{Emery+97,Bourbonnais+91}

For the pure system without $S^{\rm IFS}$ and $S^{\rm IBS}$, a
specific interaction $V_i$ in Eq.  (\ref{S0}) can render the CDW and
SC couplings irrelevant in an intermediate region of $K$, leading to
the SLL phase found in Refs.  [\onlinecite{Emery+00,Vishwanath+01}].
The interaction must be sufficiently strong and has to include at
least nearest and next-nearest neighbors.

In the presence of disorder, the scattering off impurities has to be
taken into account. Let us first focus on the effect of impurity
forward scattering (\ref{S.IFS}) in the absence of any inter-stripe
couplings of type (\ref{S-CDW-SC}). This process then changes the SLL
phase as described by $S^0+S^V$ into the disordered stripe metal
(DSM).  Introducing replicated fields and averaging over disorder
still lead to a bilinear action with correlations
\begin{subequations}
\label{Gs}
\begin{eqnarray}
  \label{G_phi}
  \langle \Phi^a_{\bf q} \Phi^b_{-{\bf q}} \rangle 
  &=& 
  \frac{\pi\delta^{ab} }
  {\omega^2/v_J+\tilde v_N(q_\perp) q_\|^2} 
  +\frac{D_\eta\delta(\omega)}{\tilde v_N^2(q_\perp)q_\|^2},
  \\
  \label{G_theta}
  \langle \Theta^a_{\bf q} \Theta^b_{-{\bf q}} \rangle
  &=& 
  \frac{\pi\delta^{ab} \tilde v_N(q_\perp)/v_J}
  {\omega^2/v_J+\tilde v_N(q_\perp) q_\|^2} 
\end{eqnarray}
\end{subequations}
with ${\bf q} \equiv (\omega,q_\|,q_\perp)$ and upper indices $a$, $b$
as replica labels.

We now examine the relevance of CDW and SC couplings and of impurity
backward scattering (IBS) with respect to the DSM state using an RG
analysis similar to that of Ref. [\onlinecite{Giamarchi+88,Giamarchi+89}].  To
first order in $D_\xi$, $\cC_m$, and $\cJ_m$ we obtain the RG flow
equations
\begin{subequations}
\label{RGF}
\begin{eqnarray}
  \label{RGF_Dxi}
  \frac{d D_\xi}{dl}&=&(3-\Delta^{\rm IBS})D_\xi,
  \\
  \label{RGF_Deta}
  \frac{d D_\eta}{dl}&=&D_\eta,
  \\
  \label{RGF_W}
  \frac{d K}{dl}&=& - \frac{2}{\pi \alpha^2 \Lambda^3}
  K D_\xi \int_{q_\perp} \tilde v_N^{-2}(q_\perp),
  \\
  \frac{d \cC_m}{dl}&=&(2-\Delta^{\rm CDW}_m)\cC_m,
  \\
  \frac{d \cJ_m}{dl}&=&(2-\Delta^{\rm SC}_m)\cJ_m,
\end{eqnarray}
\end{subequations}
with $\int_{q_\perp}\equiv \int_{-\pi}^\pi \frac{dq_\perp}{2\pi}$;
$\tilde{v}_N$ is not renormalized to first order.  The scaling
dimensions are
\begin{subequations}
\label{scal-dim}
\begin{eqnarray}
  \Delta^{\rm IBS}\!\!&=& 
  \!\!\int_{q_\perp} \tilde K(q_\perp),
  \\
  \Delta^{\rm CDW}_m\!\!&=&\!\!\int_{q_\perp}\![1-\cos(mq_\perp)]
  \left[\tilde K(q_\perp)
    + \frac{2D_\eta}{\pi\Lambda\tilde v_N^2(q_{\perp})}
  \right],\quad\quad
  \\
  \Delta^{\rm SC}_m\!\!&=&\!\!\int_{q_\perp}
  \!\!\![1-\cos(mq_\perp)] \tilde K^{-1}(q_\perp),
\end{eqnarray}
\end{subequations}
with longitudinal momentum cutoff $\Lambda \sim 1/\alpha$.  

\section{Phase Diagrams}

Before we analyze specific models for the inter-stripe interaction, we
discuss the general picture emerging from renormalization group.

\subsection{General Topology}

In the {\em absence of disorder} ($D_\xi=D_\eta=0$), $K$ preserves its
unrenormalized value $\sqrt{v_J/v_N}$.  Then the scaling dimensions
(\ref{scal-dim}) reproduce the expressions given in Ref.
[\onlinecite{Mukhopadhyay+01}].  For weak inter-stripe interactions
$V_{q_\perp} \ll v_N$ the system is in the SC phase for $K \gtrsim 1$,
whereas it is in the CDW phase for $K \lesssim 1$.  For a suitable
choice of a strong interaction $V_{q_\perp}$, an intermediate range
of $K$ can exist where all SC and CDW couplings are irrelevant and the
system is in the SLL phase. The stability of the SLL phase is
determined by the conditions $K>K^{\rm CDW}=2/\min_m\{ c_m^+ \}$ and
$K<K^{\rm SC}= \min_m\{ c_m^- \}/2$.  Hereby we define
\begin{equation}
  \label{def_cs}
  c_m^\pm \equiv \int_{q_\perp} [1-\cos m q_\perp]
  \left(\frac{v_N}{\tilde v_N}\right)^{\pm 1/2}.
\end{equation}

In the {\em presence of disorder}, the strength of impurity forward
scattering $D_\eta$ increases exponentially under the RG flow. This
has two important consequences.  First, the CDW order along the
stripes becomes now short ranged as can be easily seen from the second
term in Eq. (\ref{G_phi}). Second, it implies an exponential increase
of $\Delta^{\rm CDW}_m$ for all $m$, i.e., the \textit{irrelevance of
  weak CDW couplings}.  Thus, impurity scattering transforms the SLL
and CDW phases of the pure system into different phases.  If impurity
backward scattering (IBS) is irrelevant -- this is the case in the
entire stability region of the SLL\cite{Mukhopadhyay+01} -- a novel
phase is present which we call DSM phase.  Unlike for the SLL, the
stability of the DSM for small $K$ is no longer limited by the CDW
couplings but by IBS.  Its phase boundary is determined by the
relevance of SC couplings at large $K$ and the relevance of IBS at
small $K$.  IBS leads to {\em localization} for a bare IBS strength
$D_\xi$ larger than a critical value $D_{\xi,c}$ (that depends on
intra- and inter-stripe interactions). In this case $D_\xi$ diverges
and $K$ goes to zero under renormalization.  For $D_\xi<D_{\xi,c}$,
the system is \textit{delocalized}, $D_\xi \to 0$ and $K$ saturates at
a finite value $K^*=K^*(D_\xi)$.  For $K$ below the critical value
$K_c=3/c_\infty^+$, infinitesimal disorder produces localization
($D_{\xi,c}=0$).  For $K>K_c$, the system remains delocalized at
finite disorder strength, $0<D_\xi<D_{\xi,c}$.

To determine the phase boundary $D_{\xi,c}$ of the localization
transition for $K>K_c$, we integrate the flow Eqs.  (\ref{RGF_Dxi})
and (\ref{RGF_W}). In terms of the dimensionless disorder strength
${\cal D}=D_\xi/(\pi^3 \Lambda v_N^2)$ we obtain
\begin{equation}
\label{D(K)}
  {\cal D}(K_l)={\cal D}_0+
  \frac{c_\infty^+}{c}(K_l-K_0)-\frac{3}{c} \ln\frac{K_l}{K_0}
\end{equation}
with
\begin{equation}
  c \equiv \int_{q_\perp} \left(\frac{v_N}{\tilde v_N}\right)^2.
\end{equation}
The critical disorder strength then follows from the condition that
${\cal D}(K)=0$ at its minimum at $K=3/c_\infty^+=K_c$:
\begin{equation}
  \label{D_xi_c}
  {\cal D}_{0,c} = \frac{c_\infty^+}{c}(K_0-K_c)+
\frac{3}{c} \ln\frac{K_c}{K_0}.
\end{equation}
In the delocalized phase, the renormalized value $K^*$ of $K$ can be
obtained from Eq. (\ref{D(K)}) with ${\cal D}(K^*)=0$. The phase
boundary at large $K$ between DSM and SC phase is given by the
condition $K^*<K^{\rm SC}= \min_m\{ c_m^- \}/2$, and the boundary at
small $K$ between the DSM and the localized phase is described by Eq.
(\ref{D_xi_c}).

The actual form of the phase diagram and, more importantly, the
stability range of the SLL or DSM phase depends on the interaction
$V_{q_\perp}$ under consideration. To be specific, we will consider in
the following two models for this interaction.

%
\subsection{Model A} %
%

\begin{figure}
   \includegraphics[width=0.8\linewidth]{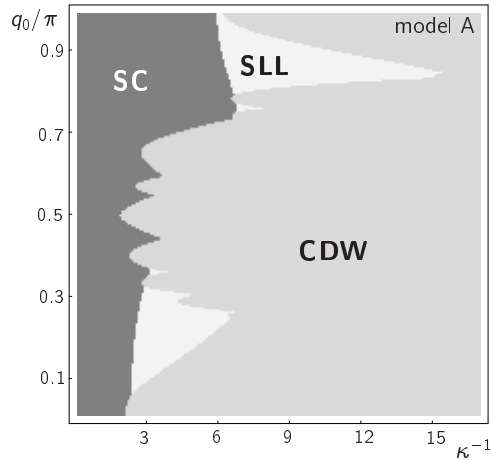}
   \caption{Phase diagram of model A for $\Delta=10^{-3}$ in the absence
     of disorder.  For large $K$, the system forms a 2D
     superconductor, while it is a 2D CDW for small $\kappa$.  The
     intermediate SLL exists only for suitable values of the parameter
     $q_0$.}
   \label{fig1}
\end{figure}

A minimal model that renders simultaneously all SC- and CDW-couplings
irrelevant for some parameter region was suggested by Vishwanath {\em
  et al.}\cite{Vishwanath+01} It assumes an inter-stripe interaction
forward scattering leading to 
\begin{equation}
  \label{Kq-A}
  \tilde
  K(q_\perp)=\kappa /[1+\lambda_1 \cos(q_\perp)+\lambda_2 \cos(2q_\perp)]
\end{equation}
Within this model, the three parameters $\kappa$, $\lambda_1$ and
$\lambda_2$ implicitly determine the intra- and inter-stripe
interaction.  The corresponding Luttinger parameter is given by
$K^2=\kappa^2/(1+(\lambda_1^2+\lambda_2^2)/2)$, and the inter-stripe
potential $V_i$ has a range of four stripe spacings. In analogy to
Ref. \onlinecite{Mukhopadhyay+01}, we rewrite $\lambda_1$ and
$\lambda_2$ as 
\begin{subequations}
\begin{eqnarray}
  \lambda_1
  &=&-\frac{4(1-\Delta)\cos q_0}{1+2\cos^2q_0} ,
  \\
  \lambda_2
  &=&\frac{1-\Delta}{1+2\cos^2q_0},
\end{eqnarray}
\end{subequations}
such that $\tilde K(q_\perp)$ takes its maximal value $\kappa /\Delta$
at $q_\perp=q_0$.  By adjusting $\Delta$ to small values, CDW
couplings are suppressed, while small $\kappa$ gives negative scaling
dimensions for SC-couplings, as pointed out by Mukhopadhyay {\em et
  al.}\cite{Mukhopadhyay+01} One thus indeed finds regions in
$(q_0,\kappa,\Delta)$-phase space, where the system is stable against all
inter-stripe couplings of type (\ref{S-CDW-SC}) and is thus in the SLL
phase.  This is demonstrated in Fig. \ref{fig1}, which may be compared
to similar plots in Refs.
[\onlinecite{Vishwanath+01,Mukhopadhyay+01}].
%
%
For sufficiently small $\Delta$, windows of $q_0$ exist where the
system evolves from a phase coherent SC through the 2D metallic SLL to
a charge ordered CDW state with decreasing  parameter $\kappa$.
Here and in the following, when both SC and the CDW couplings compete,
the one that is most strongly relevant is assumed to determine the
phase.  The actual boundary between two such strong coupling phases
might differ within a narrow corridor.  Note however, that boundaries
to either the DSM or SLL phase are obtained also quantitatively
correctly.
%
\begin{figure}
  \includegraphics[width=0.8\linewidth]{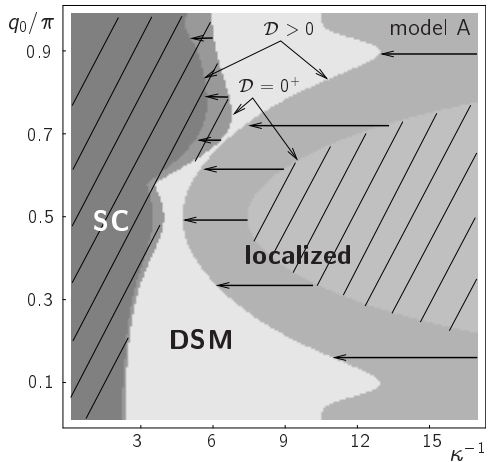}
   \caption{Phase diagram for model A with $\Delta=10^{-3}$ 
     in the presence of disorder for two disorder values.  The
     comparison of the phase boundaries at infinitesimal disorder
     $\mathcal{D}=0^+$ and finite disorder $\mathcal{D}=5 \cdot
     10^{-11}$ reflects the disorder induced renormalization of
     $\kappa$.}
   \label{fig2}
\end{figure}
%

Now disorder is added while all other parameters are unchanged. The
CDW and the SLL phases of the pure system become indistinguishable and
merge to the metallic, short-range CDW-ordered 'disordered stripe
metal' (DSM). Backscattering off impurities leads to localization in a
large portion of the former CDW phase. The SC phase shrinks through
downward renormalization of $\kappa$ by disorder (note that $\kappa$
and $K$ differ by a factor that is not renormalized).  In Fig.
\ref{fig2}, the boundaries between the three phases are given both for
infinitesimal and for finite disorder. The latter shifts the
boundaries to larger $\kappa$.

\begin{figure}
   \includegraphics[width=0.95\linewidth]{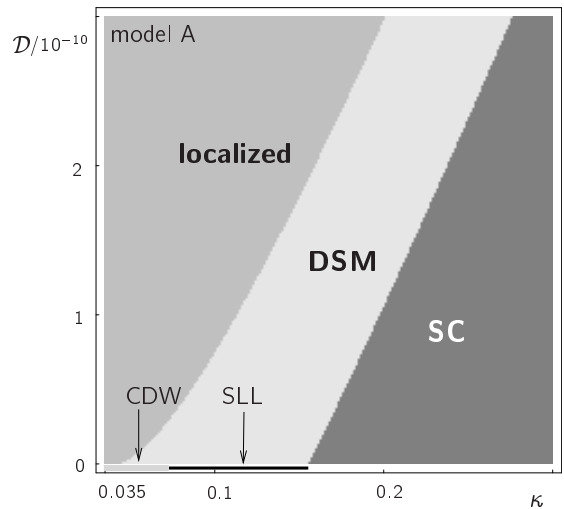}
   \caption{Phase diagram in the plane spanned by disorder and
     interaction parameter $\kappa$ for fixed inter-stripe
     interactions with $q_0=0.85\pi$ and $\Delta=10^{-3}$.  With
     increasing disorder strength $\mathcal{D}$, the boundaries of the
     DSM phase move to larger $\kappa$ keeping a nearly constant
     distance.}
   \label{fig3}
\end{figure}

A cut through Figs.~\ref{fig1} and \ref{fig2} at fixed $q_0$ but with
varying disorder is shown in Fig.~\ref{fig3}.  The SLL and CDW phases
exist only for $\mathcal{D}=0$. A delocalized phase can exist -- due
to inter-stripe forward scattering -- even for purely repulsive
interactions (for example, $q_0=0.85 \pi$, $\Delta=10^{-3}$ and
$\kappa \lesssim 1.42$ corresponds to repulsive on-stripe \textit{and}
repulsive inter-stripe interactions), as opposed to the strictly one
dimensional electron gas with delocalization for $K>3$ corresponding
to strongly attractive interactions.  However, the inter-stripe
interactions corresponding to the values of $\Delta, \kappa, q_0$
where the SLL or DSM exist may not be very realistic because of their
strength.

%
%

%
\subsection{Model B} %
%
Model A is constructed specifically in a way such that the
inter-stripe forward scattering interactions (\ref{S0}) give rise to a
nonmonotonous $\tilde K(q_\perp)$ which allows for the simultaneous
irrelevance of CDW and SC couplings in the absence of disorder.  For a
large range of parameters $q_0$ and $\Delta$, this potential has
oscillatory character in real space, which also may not be very
realistic.

A physically motivated choice for a potential that is monotonous both
in real and Fourier space may be the
screened Coulomb potential
\begin{eqnarray}
  V(r)
  &=&
  \frac{A}{r} e^{-\mu r}
\end{eqnarray}
which we consider as model B. In Fourier space this model reads
\begin{eqnarray}
  V_{q_\perp}
  &=&
  -A\ln\left[1+e^{-2\mu} -2 e^{-\mu} \cos q_\perp\right].
\end{eqnarray}
Due to the stability condition $V_{q_\perp}/v_N > -1$, see Eq.
(\ref{Ktilde}), there is a critical amplitude $A_c(\mu)$, above which
the model breaks down.

\begin{figure}
  \includegraphics[width=0.8\linewidth]{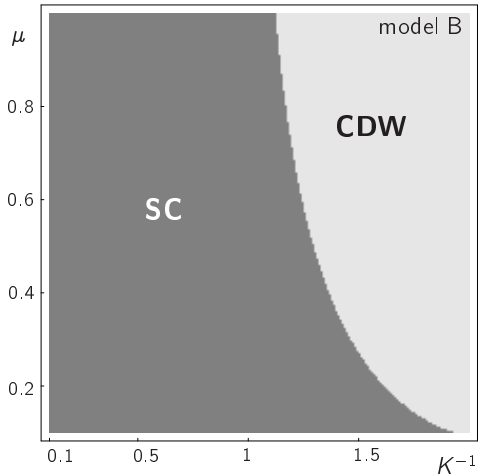}
   \caption{Phase diagram for model B for $A=0.78 v_N$ without disorder.}
   \label{fig4}
\end{figure}

\begin{figure}
   \includegraphics[width=0.8\linewidth]{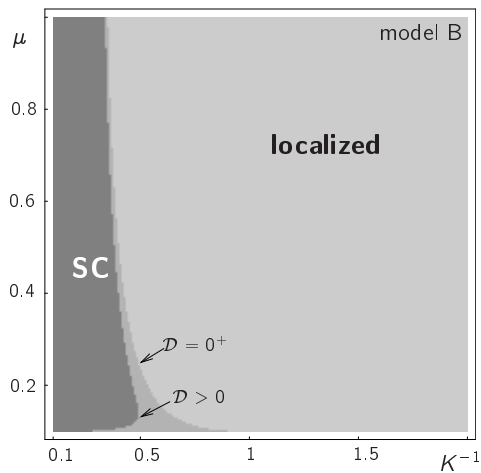}
   \caption{Phase diagram for model B 
     for two disorder values $\mathcal{D}=0^+$ and finite disorder
     $\mathcal{D}=10^{-3}$ and infinitesimal Josephson coupling
     $\mathcal{J}=0^+$; $A=0.78 v_N$;}
   \label{fig5}
\end{figure}

Fig.  \ref{fig4} displays the stability of the model with respect to
weak inter-stripe CDW and SC couplings in the absence of disorder.  No
SLL is found, the system shows for all $\mu$ and $A$ a direct
transition from the SC to the CDW phase for decreasing $K$.
%
%
The addition of disorder leads to the phase diagram in Fig.
\ref{fig5}.  As opposed to model A, impurity backscattering completely
covers the CDW phase and thus leaves only two phases, the localized
one and the SC phase.  In contrast to the competition between CDW and
SC couplings, the localization boundary is not given by the most
relevant bare coupling.  Since even weakly relevant IBS renormalizes
$K$ to small values, the disorder scaling dimension decreases while
the Josephson coupling, provided a small enough bare value, ultimately
becomes irrelevant, see Eqs. (\ref{RGF}).  Hence the boundary is given
by the onset of relevance of IBS with respect to the pure system. It
moves to larger $K$ for increasing disorder. No DSM phase is found
now.  Formally, the absence of a minimum of $V_{q_\perp}$ inside the
interval $(0,\pi)$ makes up for this latter qualitative difference in
models A and B.

%


\subsection{Correlations} %

Having established the generic topology of phase diagrams, we now
address the nature of the possible phases.  First we consider the {\em
  delocalized DSM phase}. It is described by the correlations
(\ref{Gs}) with the bare IFS amplitude $D_\eta$ and the renormalized
effective $K^*$ (in general, one has to use the renormalized but
unrescaled quantities).  We find a linear growth of the fluctuations
of $\Phi$ with longitudinal system size $L$, $\overline{\langle
  \Phi_j^2(x,\tau) \rangle}_L = (c/2\pi^2) v_N^{-2} D_\eta L$, which
leads to short-ranged longitudinal CDW correlations like for single
LLs. On the other hand, IFS does not affect the quasi long-ranged
longitudinal superconducting order (fluctuations of $\Theta$) of the
pure system.  Equally, the conductivity along the stripes is not
affected by IFS since $\eta_j(x)$ is time
independent.\cite{Giamarchi+88} From a linear-response calculation we
obtain the LL-like conductivity (which also determines the conductance
\cite{Fisher+81})
\begin{eqnarray}
  \sigma(\omega,q_\parallel,q_\perp) =  \frac {2e^2}{\pi \hbar}
  \frac{ i \omega}
  { (\omega+i0^+)^2 / v_J^* - 
  \tilde v_N(q_\perp) q_\parallel^2}  ,
\end{eqnarray}
representing a longitudinal metal. Here, $\omega$ represents a real
frequency in contrast to Matsubara frequencies in Eqs. (\ref{Gs}).
Notice that $v_J^*=v_N {K^*}^2$.  Since $K^*$ jumps from a finite
value to zero at the localization transition, $\sigma({\bf q})$
behaves discontinuously there.  In the transverse direction, CDW and
SC correlations will be short-ranged since the corresponding couplings
are irrelevant.  In the presence of a spin gap (which suppresses
single particle hopping) the irrelevance of the couplings also signals
that the DSM is a transverse insulator.

The {\em localized phase} is less amenable to an analytic description
since the divergence of IBS would necessitate a strong-coupling
analysis.  However, if the SC coupling is irrelevant, the localization
transition and the localized phase share the qualitative properties of
their 1D counterparts.  The inter-stripe couplings will lead to merely
quantitative renormalization effects.  Whether the localized phase is
a random antiferromagnet or a pinned CDW depends on the mechanism
generating the spin gap.  The longitudinal localization length $L_{\rm
  loc}$ can be estimated analytically when the transition line is
approached from the localized side. For $K<K_c$, $L_{\rm loc} \simeq
\Lambda^{-1} {\cal D}_0^{-1/(3(1-K/K_c))}$.  For $K>K_c$, $L_{\rm loc}
\simeq \Lambda^{-1} \exp(\tilde c /\sqrt{c ({\cal D}_0-{\cal
    D}_{0,c})}\,)$ with $\tilde c$ a numerical factor of order unity.
Thus, the inter-stripe interactions influence the localization length
quantitatively, but the qualitative behavior found\cite{Giamarchi+88}
for the localization transition in a single LL persists.

\section{Discussion and Summary}

In summary, we have examined impurity effects in arrays of coupled
LLs.  The competition between impurity backscattering, CDW and SC
couplings allows for three different phases: a localized phase,
a superconducting phase, and the disordered stripe metal.  The latter
two phases are delocalized since IBS scattering is irrelevant.  While
for a single stripe delocalization occurs only for strongly
attractive on-stripe interactions ($K \gtrsim 3$), for a coupled
stripe array delocalization is possible also for purely
\textit{repulsive} on-stripe and inter-stripe interactions (forward
scattering). 

The delocalized DSM phase is metallic in longitudinal direction and
insulating in transverse directions. Its correlations for CDW order
are short-ranged in all directions whereas superconducting
correlations are quasi long-ranged along the stripes and short-ranged
in the transversal direction.  These experimentally accessible
features should allow to identify the disordered stripe metal and to
distinguish it from the SLL phase.

In the above analysis we have determined the phase diagram from a
stability analysis of a Gaussian fixed point -- representing the
stripe array with forward scattering by interactions and impurities
-- with respect to CDW and SC couplings as well as impurity backward
scattering.  This stability analysis, reflected by the flow equations
(\ref{RGF}) which are linear in $D_\xi$, $\cC_m$, and $\cJ_m$, requires the
weakness of these couplings.  In principle, this approach
does not cover strong-coupling phenomena: sufficiently strong
couplings might drive transitions into SC or CDW phases which are less
susceptible to disorder than the SLL.

Although we cannot consistently access such strong coupling phenomena
via our flow equations, they nevertheless can be used to determine
crossovers related to the \textit{relative} strength of non-Gaussian
couplings.  Since all CDW couplings are irrelevant at the DSM fixed
point, we raise the question of whether a CDW phase can be
reestablished if CDW couplings are sufficiently strong in comparison
to disorder.  We focus on the region where a CDW coupling $\cC_m$ is
relevant in the absence of disorder and where IBS is irrelevant (i.e.,
$\delta \equiv 2- \Delta^{\rm CDW}_m >0 $ and $\Delta^{\rm IBS}>3$
for the bare parameters).

Irrespectively of the relative strength of CDW couplings and IBS, the
presence of disorder implies a continuous growth of $D_\eta$ and thus
also of $\Delta^{\rm CDW}_m$ which implies the irrelevance of $\cC_m$
only on sufficiently large scales.  Thus the question is, whether
$D_\eta$ increases fast enough to achieve $\Delta^{\rm CDW}_m>2$
before a strong-CDW coupling regime is entered.  This regime is
entered when the dimensionless coupling $\hat \cC_m \equiv
\mathcal{C}_1/(\Lambda^2 v_N)$ becomes of order unity under
renormalization before the disorder contribution to $\Delta^{\rm
  CDW}_m$ becomes of order $\delta$.  For weak $V_{q_\perp}$ or $m \gg
1$, this is the case if
\begin{eqnarray}
  \hat \cC_m \gtrsim  \left( \frac{2 c D_\eta}{\pi \Lambda v_N^2
      \delta} \right)^\delta.
\end{eqnarray}
For CDWs (quasi-)long-ranged charge correlations can exist in the
presence of disorder only in $D>2$ dimensions like for vortex
lattices.\cite{Scheidl+00} Then the fermions would form a pinned
(localized) Wigner crystal.  However, in $D=2$, the formation of CDW
order is prohibited by the proliferation of
dislocations,\cite{Villain+84} which ultimately render the CDW
coupling irrelevant on sufficiently large scales.

\begin{acknowledgments}
  We would like to thank T. Nattermann and B. Rosenow for comments and
  in particular T. Giamarchi for discussions on ambiguities of
  the renormalization procedure in an early version of the manuscript.
  This work was supported by the Deutsche Forschungsgemeinschaft
  through the Emmy-Noether grant No. EM70/2-1 (SB and TE).
\end{acknowledgments}

\vspace*{-.5cm}


\end{document}